\documentclass{jpp}

\usepackage{epsf}

\title[Riga dynamo experiments]{Self-excitation in 
a helical liquid metal flow:
The Riga dynamo experiments}

\author[A. Gailitis et al.]%
{A. Gailitis$^1$, G. Gerbeth$^2$, Th. Gundrum$^2$, O. Lielausis$^1$, 
G. Lipsbergs$^1$, 
E. Platacis$^1$,
and F. Stefani$^2$}

\affiliation{$^1$Institute of Physics, Latvian University,
LV-2169 Salaspils 1, Riga, Latvia\\[\affilskip]
$^2$Helmholtz-Zentrum Dresden-Rossendorf,
Bautzner Landstr. 400, D-01318 Dresden, Germany}

\pubyear{2006}
\volume{???}
\pagerange{???--???}
\date{?? and in revised form ??}
\begin{document}

\maketitle

\begin{abstract}
The homogeneous dynamo effect is at the root of magnetic
field generation in cosmic bodies, including planets,
stars and galaxies. While the underlying theory had 
increasingly flourished since the 
middle of the 20th century, hydromagnetic 
dynamos were not realized in laboratory until 1999. 
On 11 November 1999,
this situation changed with the first 
observation of a kinematic
dynamo in the Riga experiment. Since that time,
a series of experimental campaigns has provided a wealth of
data on the kinematic and the saturated regime.
This paper is intended to give a comprehensive survey
about these experiments, to summarize their main results 
and to compare them with numerical simulations.
\end{abstract}

\section{Introduction}

The seminal paper of Steenbeck, Krause \& R\"adler (1966), 
which is celebrated in this special issue, was not only a landmark
in the theoretical description of cosmic dynamos, but had also
fostered a series of experimental activities.
One of them was the experimental demonstration  
of the $\alpha$--effect in the Riga ``$\alpha$--box,''    
a system of two orthogonally 
interlaced copper channels (Steenbeck {\it et al.} 1967).             
Since the
sodium flow through this system was not mirror-symmetric,
it produced a measurable $\alpha$--effect, as confirmed
by the observations that the induced voltage
in weak fields was proportional to $B v^2$ while 
it changed sign when the applied magnetic field was 
reversed. 

In the same year, the first author of this paper (Gailitis 1967)
proposed an experimental dynamo 
``in which the gyrotropic turbulence is simulated by means
of a certain pseudo-turbulent motion.'' This early idea 
to substitute 
real helical (``gyrotropic'') turbulence 
by ``pseudo-turbulence'' was later to be realized 
in the two-scale Karlsruhe dynamo experiment 
by using 52  parallel channels with helical sodium flows 
inside (M\"uller \& Stieglitz 2000; Stieglitz \& M\"uller 2001;
M\"uller, Stieglitz \& Horanyi 2004; 
M\"uller, Stieglitz \& Horanyi 2004).

Meanwhile, the focus in Riga 
had shifted away from the 
two-scale, multi-channel flow dynamo towards 
a dynamo concept based on a single helical flow. 
This development was ignited by the
paper of Ponomarenko (1973) who had proved 
dynamo action for a solid
conducting
rod screwing through a medium of infinite extend
with which it is in sliding electrical contact.
 
In a detailed numerical analysis of this 
''elementary cell'' of a dynamo,
Gailitis and Freibergs (1976) had 
computed a remarkable low
critical magnetic Reynolds number of 17.7 for 
the convective instability. Later it was found
that by adding a concentric straight 
backflow to the 
helical flow this
convective instability could be rendered 
an absolute one (Gailitis and Freibergs 1980).

\begin{figure}
\begin{center}
\epsfxsize=13.4cm\epsfbox{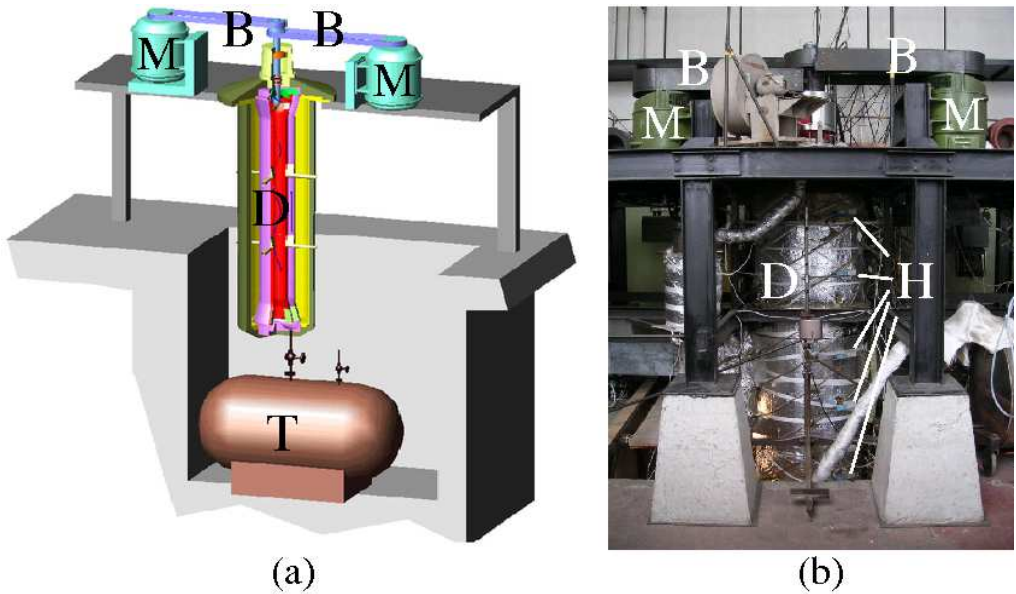}
\end{center}
\caption{Sketch (a) and photograph (b) of the Riga dynamo facility, 
with the central dynamo 
module (D) mounted on a
steel frame, the two
DC-motors (M), the belts (B),  
one of the storage tanks (T), and the external Hall sensors (H).}
\label{fig1}
\end{figure}

The Riga dynamo experiment (figure \ref{fig1})
is the laboratory realization of
this principle of magnetic field self-excitation
in a single helical flow. 
At this facility, an exponentially growing eigenmode
was observed for the first time in November 1999
(Gailitis {\it et al.} 2000a).
A follow-up experiment, reaching also the saturation regime,
was carried out
in July 2000 and reported in a number of papers
(Gailitis {\it et al.} 2001a; Gailitis {\it et al.} 2001b).

Since those days, a number of 
further experimental campaigns 
have been
carried out: the experiments in 
June 2002, February 2003,
July 2003, May 2004, February/March 2005, and July 2007
have brought about
many details on the spatial and temporal magnetic field
structure, due to a step-by-step improvement of the
measuring system. The wealth of new data has led to an 
improved understanding
of the experiment, which we would like to report on 
in this survey. Their correspondence 
with numerical simulations will be another
focus of this paper.

After three further campaigns in  April 2009 and
February and June 2010, which were less successful due to 
several technical problems, 
the Riga dynamo was disassembled and completely 
reconstructed,
before two new campaigns were carried out 
in June 2016 and
April 2017. Their results, however, 
will be discussed elsewhere.

\section{Theory and numerics}

Dynamo theory is governed by two coupled equations. The first
one is the induction equation
for the magnetic field $\bf B$,
\begin{eqnarray}
\frac{\partial {{\bf{B}}}}{\partial t}=\nabla
\times ({\bf{v}} \times {\bf{B}})
+\frac{1}{\mu_0 \sigma} \Delta {\bf{B}} \; ,
\label{eq1}
\end{eqnarray}
where $\bf v$ denotes the velocity field of the fluid, $\sigma$
its electrical conductivity and $\mu_0$ the magnetic 
permeability of the
free space. Equation (\ref{eq1}) follows directly from
Amp\`{e}re's, Faraday's and Ohm's law. In its derivation, 
quasi-stationarity is assumed in the sense that the 
displacement current can be neglected.

Evidently, without any flow velocity, 
(\ref{eq1}) reduces to a diffusion
equation which describes
the free decay of the magnetic field in an 
electrically conducting medium.
The velocity dependent term can,
under certain conditions
for the strength and topology of the flow, 
counteract this decay
and  lead to a positive gain for the magnetic field.

As long as the velocity is stationary, (\ref{eq1}) can be 
transformed into
an eigenvalue equation
\begin{eqnarray}
\lambda {\bf{B}}=\nabla
\times ({\bf{v}} \times {\bf{B}})
+\frac{1}{\mu_0 \sigma} \Delta {\bf{B}} \; ,
\label{eq2}
\end{eqnarray}
with $\lambda=p+2 \pi i f$, including the growth rate $p$ and the
frequency $f$. It is rather 
typical for dynamos that they are governed by
non-self-adjoint induction operators, which may have, 
in general, complex eigenvalues.

In case of a positive growth rate, the magnetic field
will increase exponentially until
it reaches such an amplitude that the back-reaction of 
the Lorentz forces on the
velocity is not negligible anymore.
Then it becomes necessary to consider also the 
Navier-Stokes equation
\begin{eqnarray}
\frac{\partial {{\bf{v}}}}{\partial t}+({\bf{v}}
\cdot \nabla) {\bf{v}}&=&
- \frac{\nabla p}{\rho}+ \frac{1}{\mu_0 \rho}
(\nabla \times {\bf{B}})
\times {\bf{B}}+\nu \Delta  {\bf{v}}+{\bf{f}}_{prop} \; ,
\label{eq3}
\end{eqnarray}
where  $\rho$ and $\nu$ denote                          
the density and the kinematic viscosity of the fluid,
and ${\bf{f}}_{prop}$ symbolizes the
force exerted by the propeller (in our context). 
Typically, 
the dynamo will saturate into a state where 
the Lorentz forces act against the source of its
generation (Lenz's rule).

\subsection{Kinematic regime of the Riga dynamo}

The basic idea of the Riga dynamo experiment traces 
back to the paper of Ponomarenko (1973) who had proved 
that dynamo action can occur when a conducting rod 
of infinite length screws
slidingly through a conducting medium of infinite 
radial extension.

In a more detailed numerical analysis of this 
''elementary cell'' of a dynamo,
Gailitis and Freibergs (1976) had found a 
remarkable low
critical magnetic Reynolds number of 17.7
for the convective instability.
By adding a back-flow, this convective instability can be
made into an absolute instability 
(Gailitis \& Freibergs 1980). All these early
computations, including the main geometric 
optimization of the three 
cylinders configuration (Gailitis 1996), were 
carried out with a one-dimensional code in which 
the induction equation  (\ref{eq1})                                     
was reduced to three coupled radial equations for 
the components of 
$\tilde{\bf B}(r,\lambda)$, defined via 
${\bf{B}}(r,\phi, z,t)=\exp{(im \phi+ikz+\lambda t)}
\tilde{\bf B}(r,\lambda)$. Such a  simplification is
well justified in case that the velocity is 
independent of $\phi$, $z$ and $t$.
The method is not restricted to study convective 
instabilities; using spatially growing modes it can also 
be employed to identify absolute instabilities by 
searching for saddle-points in the complex plane which
indicate zero group velocity.

\begin{figure}
\begin{center}
\epsfxsize=13.4cm\epsfbox{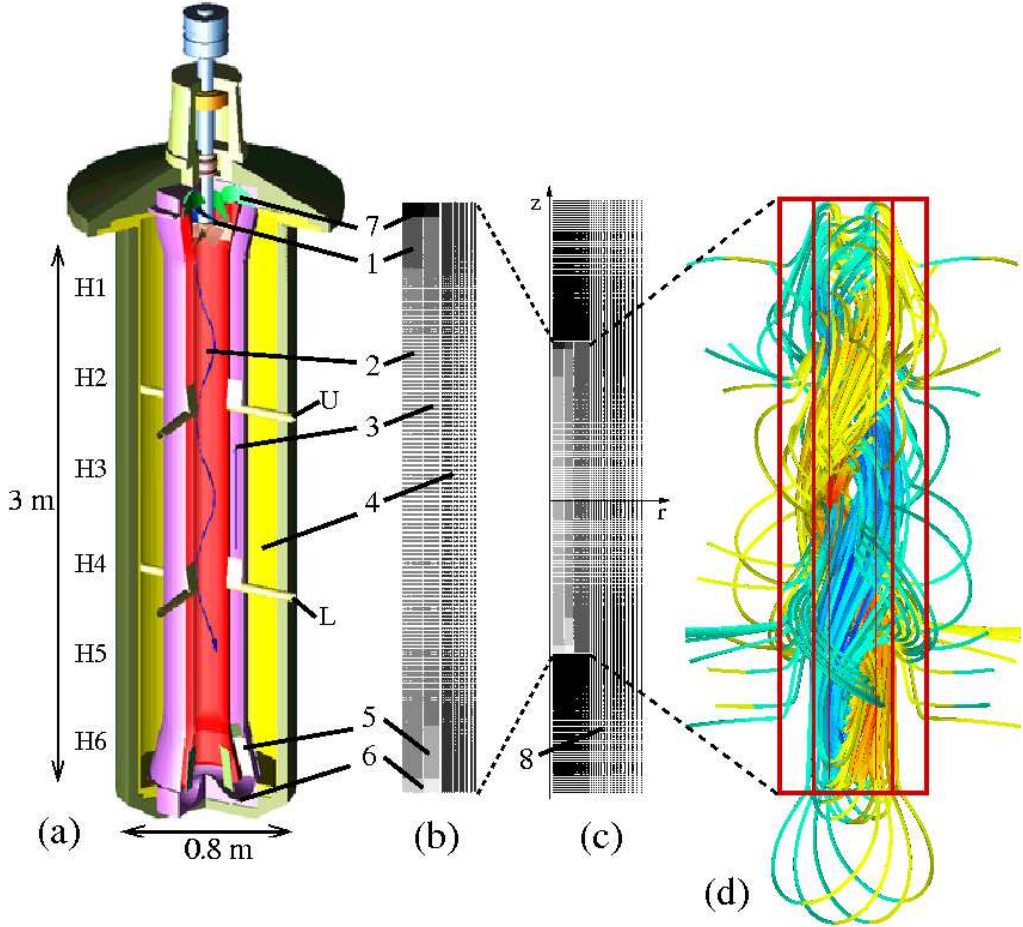}
\end{center}
\caption{(a) Central module of the Riga dynamo experiment, 
(b,c) geometrical simplification for 2D 
numerical simulations,  and  
(d) the structure of its eigenfield.
1 - Propeller region; 2 - Helical flow region;
3 - Back-flow region; 4 - Outer sodium filled cylinder;
5 - Straightening region behind the lower bending;
6 - Lower bending; 7 - Upper bending; U - Upper port; L - Lower Port; 
H1...H6 - External Hall sensors. 
(c) Embedding of the dynamo in vacuum.
The grid number is 401 in axial direction and 64 in radial direction.
Note the non-uniform grid spacing with the finest resolution in the
bending regions. The snapshot of the computed magnetic field lines 
in the
kinematic regime was computed 
for a propeller rotation rate of 2000 rpm.
The field pattern rotates with a frequency 
of 1.16 Hz around the vertical axis,
in the same direction as the flow.
The color
of the field-lines indicates the z-component of the field. The red frame
indicates the simplified geometry as it is used  in the simulations.}
\label{fig2}
\end{figure}

Complementary to this 1D code, a two-dimensional 
(2D) code was developed which,
as long as the velocity is unchanged along the vertical axis, 
provided results very close to those of the 1D 
code (Stefani, Gailitis \& Gerbeth 1999; Gailitis {\it et al.} 2004). 
Its main advantage is the possibility to cope also with
velocity fields
varying along the axis of the dynamo.
This 2D code, which was used
for many simulations of the Riga dynamo, 
is a finite difference scheme on a non-homogeneous grid,
with an Adams-Bashforth method of second order for the 
time integration.
The real geometry of the dynamo module has been slightly
simplified (see figure \ref{fig2}b,c), with all curved 
parts in the 
bending regions being replaced by rectangles. The velocity 
in the central cylinder was inferred from a number of 
measurements at a water test facility
and some extrapolations 
(Christen, H\"anel \& Will 1998; 
Stefani, Gailitis \& Gerbeth 1999). 
These measurements had revealed a
slight decay of the rotational component along the flow.
The axial velocity in the back-flow region 
was assumed as 
purely axial and
constant so that the mean flow is equivalent to that 
in the inner cylinder.
In the bending regions some simplified flow structures 
were employed,
ensuring the divergence-free condition and rather smooth
transitions from the central to the back-flow cylinder and 
vice-versa.
In the propeller region the rotational component is 
assumed to increase linearly. 

Appropriate interface condition were applied at the
walls of the cylinders where velocity jumps occur. 
The three conditions used in the numerical scheme are 
that the two tangential components
of the electric field are continuous and that 
the divergence is zero.
A notorious problem for dynamos in non-spherical geometry
is the treatment of the boundary conditions
at the outer rim (Stefani, Giesecke \& Gerbeth 2009).
Our solution for this problem was as follows: in the  
dynamo domain, the Adams-Bashforth-scheme was applied. 
For every time-step,
we solved the Laplace equation for the magnetic field 
in the exterior of the domain by means of a 
pseudo-transient relaxation. 
At the outer boundary of this
extended domain we use zero boundary conditions, whereas 
at the interface to the dynamo
we use the interface conditions. Admittedly 
this is a tedious and time-consuming
procedure. However, since a simpler method of using
vertical field conditions lead to a remarkable 
underestimation of the critical magnetic Reynolds number, 
the correct treatment of the exterior pays off when it comes 
to an accurate prediction of the dynamo.

Figure \ref{fig2}d gives an impression of the magnetic field line 
structure as it results from the 2D code. 
The double-helix field 
structure
is clearly visible. Note that this is a snapshot 
of the field pattern which actually  
corotates with the flow around
the vertical direction, although with a much 
lower frequency than the rotational
velocity.

\subsection{Saturation regime of the Riga dynamo}

In contrast to the kinematic regime, the understanding of the 
saturation regime is much more
intriguing. In principle, it requires
the fully coupled three-dimensional numerical simulation of 
the induction and Navier-Stokes
equations, the latter one at a Reynolds number 
of  appr. $2 \times 10^6$. 
While related numerical efforts have indeed been
undertaken by using a RANS turbulence model  
(Kenjere\v{s} {\it et al.} 
2006; Kenjere\v{s} \& Hanjali\'{c} 2006; Kenjere\v{s} 
\&  Hanjali\'{c} 2007), we will rely in the following
on the simple back-reaction mechanism 
as described by Gailitis {\it et al.} (2002b).

In this model we focus on the dominant effect
of braking the azimuthal component. We start by splitting the
velocity field into an undisturbed
part ${\bar{\bf{v}}}$, which we assume to be a solution of
equation (\ref{eq3}) 
{\it{without the Lorentz force term}}, and
a perturbation ${\delta{\bf{v}}}$ which we assume to be
caused by the Lorentz force. The same
splitting is done for the pressure: 
$p=\bar{p}+\delta p$.

In the first order approximation we can now
compute the magnetic field ${\bf{B}}$
for a given (i.e., measured or interpolated)
undisturbed $\bar{\bf{v}}$ using the
induction equation,
and insert the resulting magnetic field into
the linearized Navier-Stokes
equation for the flow perturbation ${\delta{\bf{v}}}$,

\begin{eqnarray}
\frac{\partial {{\delta{\bf{v}}}}}{\partial t}+({\bar{\bf{v}}}
\cdot \nabla) {{\delta{\bf{v}}}}+
({\delta{\bf{v}}}
\cdot \nabla) {{\bar{\bf{v}}}}=&\frac{1}{\mu_0 \rho}
(\nabla \times {\bf{B}})
\times {\bf{B}}- \frac{\nabla \delta p}{\rho}+\nu \Delta  {\delta{\bf{v}}} \; .
\label{eq4}
\end{eqnarray}

In principle, this perturbation method can be extended 
to higher orders
by inserting the resulting $\bf{v}={\bar{\bf{v}}}+{\delta{\bf{v}}}$
again into the induction equation, and so on.

Here, we present a simplified version
which will later be shown to explain the observed
back-reaction effects.
The simplifications are as follows:
First, we restrict the perturbation method to the
first order.
Second we skip the viscous term in equation (\ref{eq4}).
Third, we restrict all back-reaction
effects to their axisymmetric contribution 
(since the azimuthal
dependence of the self-excited magnetic field is proportional to
$\exp{(im\phi)}$ with $m=1$, the Lorentz force in
Equation (\ref{eq4}) contains terms with $m=0$ and $m=2$ 
leading to corresponding velocity and pressure perturbations).
Fourth, we do not even solve the simplified equation (\ref{eq4})
but concentrate on the azimuthal component.
This simplification is motivated
by the fact that one can expect the $z$-component of the
Lorentz force
to result in an additional
pressure gradient in $z$-direction and not into
a velocity decrease
(the total flowrate $\int r v_z dr d\phi$ has to
be constant along the $z$-axis).
The remaining rotational part of the radial and the
axial force will, of course, result in some
deformation
of the velocity profile.
The only component of the $m=0$ Lorentz force part
which by no means
can be absorbed into a pressure gradient is the
azimuthal one on which we focus in the following.

After all, we solve the  ordinary differential
equation for the perturbation  $\delta v_{\phi}$,
\begin{eqnarray}
\bar{v}_z \frac{\partial}{\partial z} \delta v_{\phi} =  
\frac{1}{\mu_0 \rho} [(\nabla \times {\bf{B}}) \times
{\bf{B}}]_{\phi}  \; ,
\label{eq5}
\end{eqnarray}
both in the helical flow region and in the back-flow 
region. As shown in Gailitis {\it et al.} (2002b), 
the Lorentz force resulting from the magnetic eigenfield
illustrates Lenz's rule. Its axial component breaks the 
axial velocity in both flow regions, thereby leading to a 
pressure increase which has to be overcome by 
additional power of the motors. At the same time, the
azimuthal component leads to reduction of the 
differential rotation, by reducing the rotation in the
helical flow region and {\it accelerating} it in the back-flow
region. This way, the generation capacity of the saturated dynamo 
decreases along 
the downward flow, which finally leads to a general upward 
shift of the eigenfield.

\section{The facility and the experimental campaigns}

After, in 1987, a forerunner experiment
had shown 
significant field 
amplification but no self-excitation (Gailitis {\it et al.}  (1987), 
more than a decade was spent
on the design and the optimization
of the present machine. 
The overall structure is depicted in figure \ref{fig1}:
the central module of approximately 
3 m length and 0.8 m diameter is
mounted on a steel frame. The power for the 
propeller is provided by two DC-motors, which can be driven 
up to 200 kW.
Before and after the experimental campaigns 
the sodium resides in two storage tanks.

Basically, the dynamo module consists of three concentric
cylinders with different flow structures (figure \ref{fig2}a). 
In the central 
cylinder the sodium is forced by the propeller, and guided by
pre- and post-propeller vanes, on a helical
path with an appropriate relation of axial and azimuthal velocity
components and an optimized radial dependence of both.
After the flow swirl is taken out by some blades 
in the lower bending region, the
flow becomes basically rotation-free in the 
back-flow cylinder, although 
the back-reaction of the magnetic field
in the saturation regime will induce some rotation within this
tube. The same holds for the third, outermost, cylinder where the sodium
is stagnant at the beginning of the experiment, 
but where the Lorentz forces 
will also drive some flow when 
the magnetic field has become
strong enough. 

In  the following we will summarize the
results of the experimental campaigns carried out
between 1999 and 2007.

\subsection{November 1999}

The main result of the experimental campaign in November 1999
was the detection of a flow induced, slowly growing eigenmode
of the dynamo. Since in this first experiment 
the main focus laid on the
study of the {\it amplification} of an externally applied 
field, the detection of the 
{\it self-excited} mode was more a less a by-product.

\begin{figure}
\begin{center}
\epsfxsize=13.4cm\epsfbox{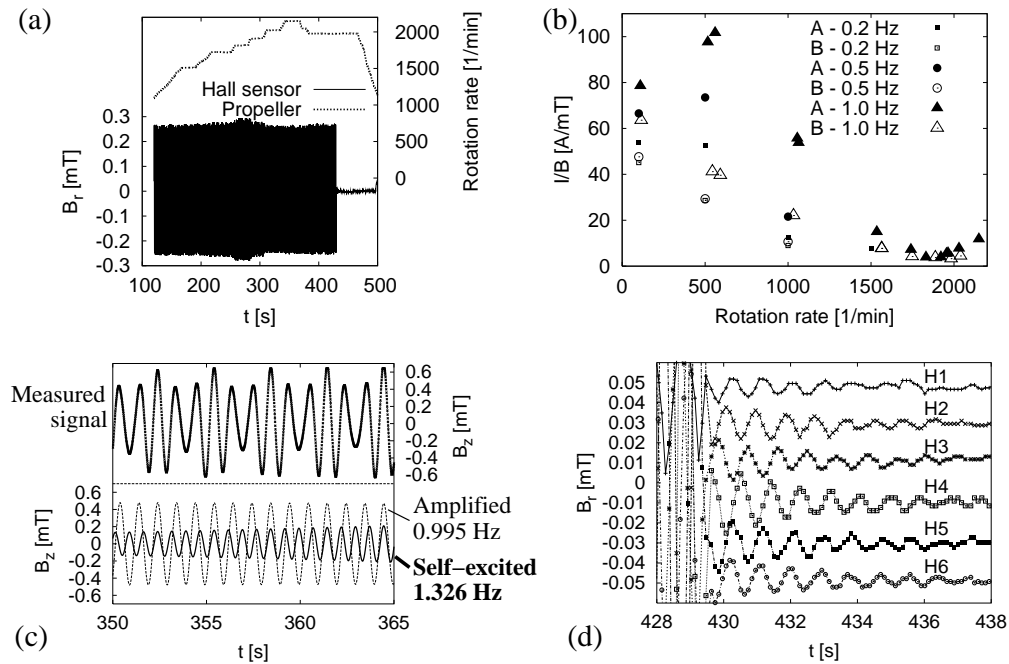}
\end{center}
\caption{The last, and most relevant, 
experiment of November 1999. (a) The dotted 
line gives the propeller rotation rate, 
the ''patch'' shows the oscillating radial field 
measured at the external 
Hall sensor H4 situated at 1.85 m 
below the upper frame. (b)
Inverse amplification for different frequencies and phase 
relations of the currents in the excitation coils. The minimum
around 1800 rpm reflects the strongest amplification, 
but does not indicate self-excitation. 
(c) For the highest rotation rate of
2150 rmp, a superposition of the amplified excitation field (0.995 Hz)
and a self-excited field (1.326 Hz) appears. (d) After switching of
the external excitation at 1980 rpm, the Hall sensors H1...H6 
show a very slow exponential decay, with typical amplitude and 
phase relations of the magnetic eigenfield.}
\label{fig3}
\end{figure}

Figure \ref{fig3}a shows the magnetic field
as measured by one of the external Hall sensors 
(H4, situated at a vertical distance of 
1.85 m from upper frame) 
over a time span of 400 seconds. During this time,
the propeller rotation rate went up from 
1000 rpm to 2150 rpm, then it remained for a while at 
1980 rpm before being reduced to zero. During this time
the signal of the external Hall sensor changes
only slightly, with a shallow maximum at approx. 
1800 rpm. What the Hall sensor sees here is essentially
the magnetic field from the external excitation 
coils which is
only slightly modified by the induction effect of
the sodium flow. The maximum at 1800 rpm 
can be understood
when showing (see figure \ref{fig3}b) the 
ratio of external coil current to the
magnetic field  
as measured by an internal fluxgate sensor. This
ratio is a measure of the {\it inverse} amplification 
of the externally  applied excitation 
field. Irrespective
of some particular dependencies on the frequency and
the phase relation of the external 3-phase current, 
this ratio shows a clear minimum at around 1800 rpm, 
with the amplification reaching a value of 20.
However impressive this factor might appear,
it testifies only a strong amplification, but not  
self-excitation! 
All the data points shown in figure \ref{fig3}b 
rely on measured magnetic field data 
with exactly the frequency of the 
external excitation.

The only exception is the rightmost point. Here,
at a rotation rate of 2150 rpm (and a still rather high
temperature of 210$^{\circ}$C), a superposition of the
amplified excitation signal (at 1 Hz) 
and a self-excited eigenmode (1.326 Hz)
was recorded for a period of 15 seconds
(the interval between 350 s and 365 s in 
figure \ref{fig3}c). 
Yet, this period was too short
to allow the field to grow to values relevant for
back-reaction effects. Soon after that measurement, the 
rotation rate was reduced to 1980 rpm,
where the excitation was then switched of appr. 
at 429 s. The Hall sensor data in 
Figure \ref{fig3}d mirror the extremely slow
decay of the magnetic field, indicating 
that the dynamo was only slightly sub-critical at 
this stage. At 470 s, the experiment had to be 
terminated
due to a technical problem with a minor sodium
leak at the bearing of the propeller shaft.

As  for the observed 
superposition of the amplified signal and the
self-excited signal shown in \ref{fig3}c 
one might argue whether this
was really self-excitation, or if there was 
some triggering 
of the eigenmode by the applied magnetic field.
Below we  will see that the growth rates and frequencies of the
exponentially increasing (figure \ref{fig3}c) 
and the slowly decreasing (figure \ref{fig3}d) signals 
fit perfectly 
into the other data from later experiments.
This is a strong argument that the 
exponential growth was indeed the first
realization of the kinematic dynamo regime in a 
liquid metal experiment.

\subsection{July 2000}

After having repaired the broken seal of the propeller shaft,
the next campaign was carried out in July 2000. It  
comprised four runs (see figure \ref{fig4}) providing
now the first results on the saturated dynamo regime.
Here and in the next figures, we show the propeller 
rotation rate and the radial magnetic field measured 
by Hall sensor H4.
While run 1  was still completely in the
subcritical regime where only amplification could be studied, 
in run 2 and run 3  the excitation current was 
completely switched of after some 100 s. The big 
''blobs'' in in the runs 2-4 indicate 
the attainment of the saturation regime, with 
the field amplitudes clearly depending on the specific 
propeller rotation rate. In Run 4, the excitation current was 
switched off from the very beginning. 

\begin{figure}
\begin{center}
\epsfxsize=13.4cm\epsfbox{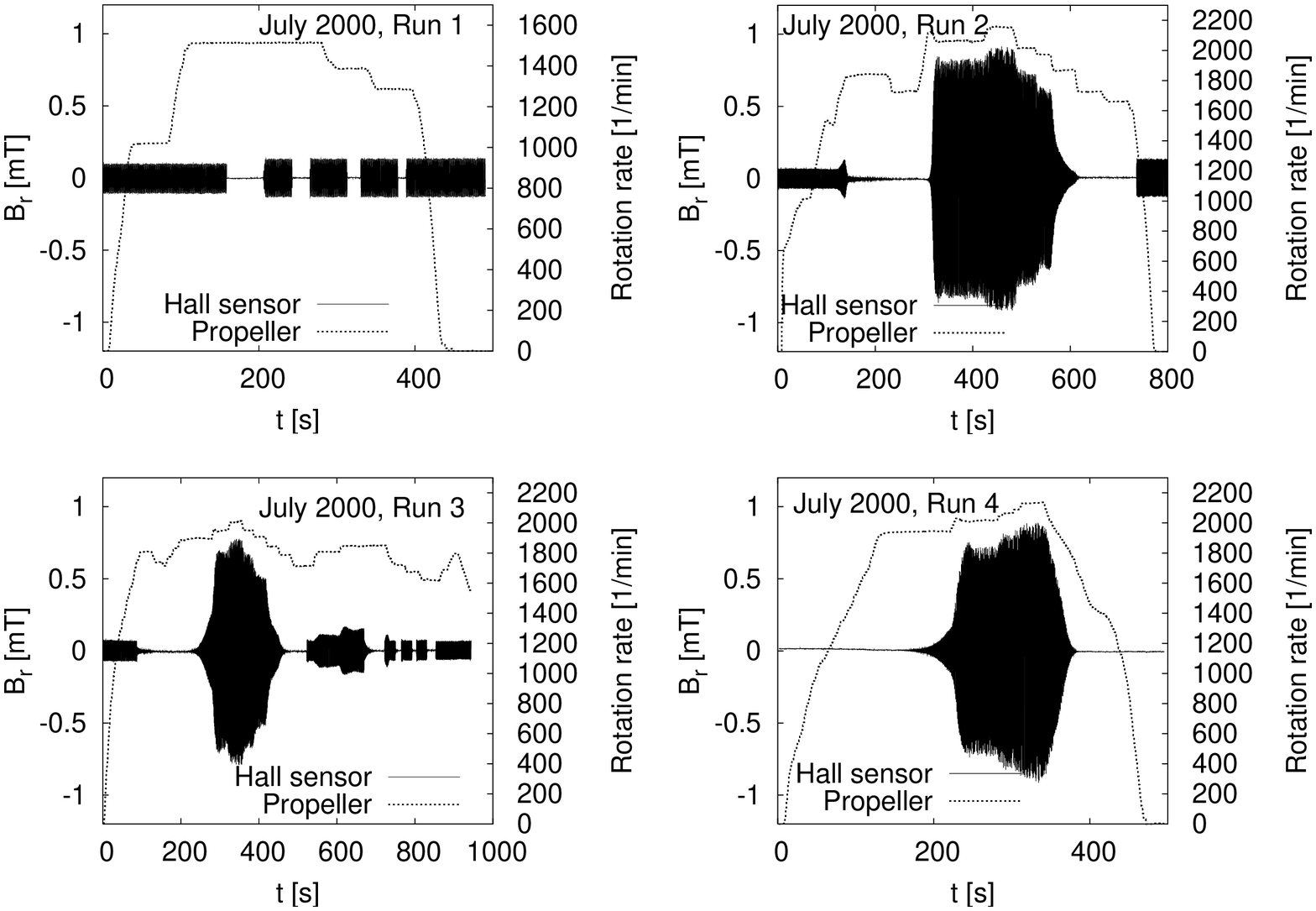}
\end{center}
\caption{Experiments in July 2000. Propeller rotation rate and
radial magnetic field measured at H4. 
While run 1 was still under-critical,
runs 2-4 show self-excitation when the rotation rate exceeds some 
critical value of appr. 1840 rpm. For run 4, the excitation coil
was switched off all the time.}
\label{fig4}
\end{figure}

\subsection{June 2002}

\begin{figure}
\begin{center}
\epsfxsize=13.4cm\epsfbox{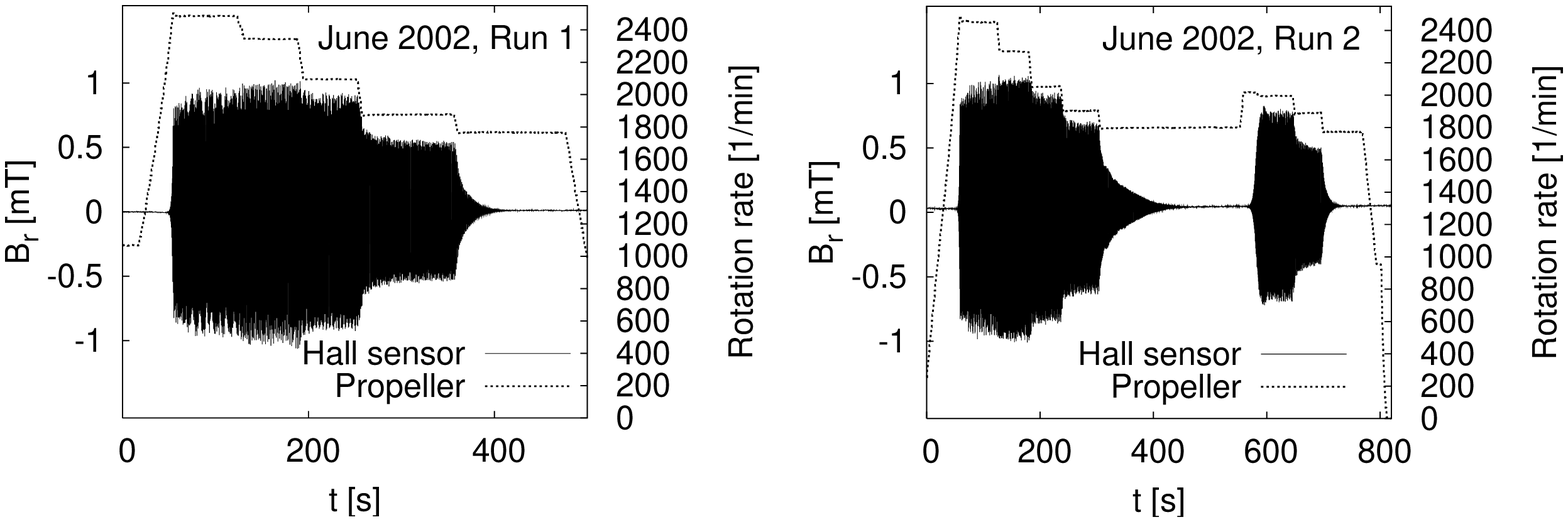}
\end{center}
\caption{Experiments in June 2002. }
\label{fig5}
\end{figure}

In June 2002, a total of eleven runs were carried out, from which
only two runs provided usable results (figure \ref{fig5}). The remaining runs were
spoiled  by problems with the control of the Argon pressure system 
which resulted
probably in an Argon inflow into the dynamo module and a non-optimal 
coupling of the propeller  speed to the fluid.
However, as will be shown below, 
the growth rates and the frequencies for a given 
rotation rate (and temperature) provide an 
accurate instrument to 
distinguish such sub-optimal runs from the
''good'' ones.

\subsection{February 2003}

\begin{figure}
\begin{center}
\epsfxsize=13.4cm\epsfbox{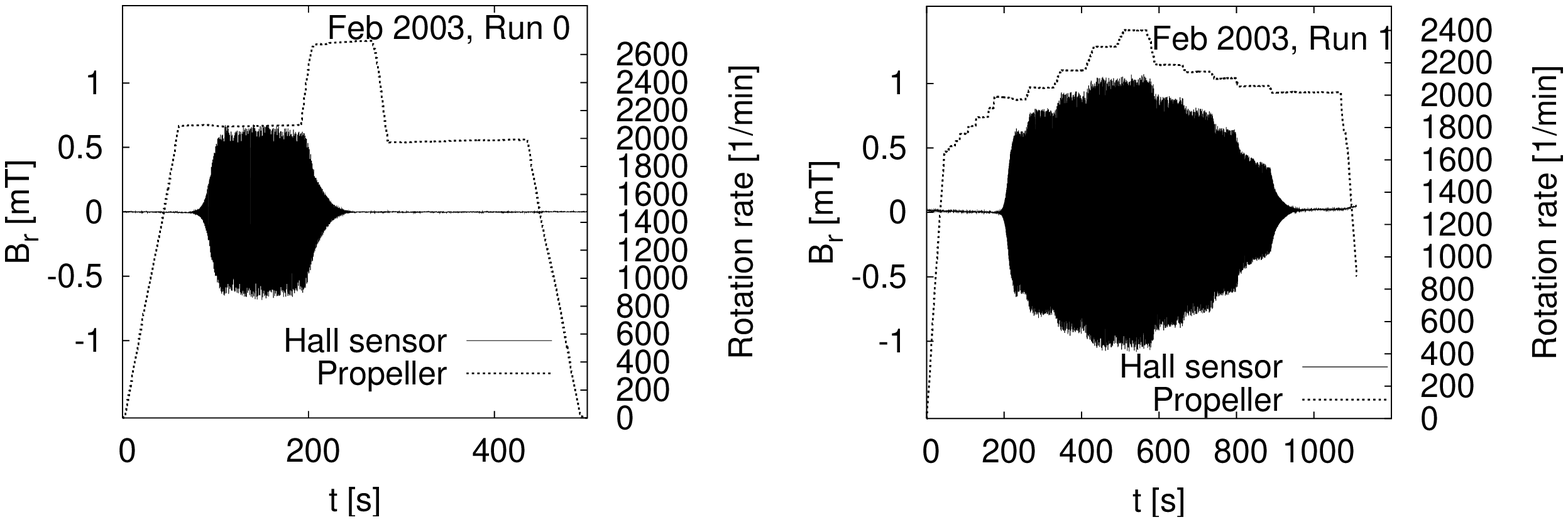}
\end{center}
\caption{Experiments in February 2003. 
Run 0 shows an abnormal behaviour
which was, very likely, due to cavitation. Run 1 was
carried out with a fine gradation of the rotation rate.}
\label{fig6}
\end{figure}

Of the February 2003 campaign, we document two 
runs in figure \ref{fig6}. Obviously, run 0 was not 
completely successful, since after the dynamo had started
at a rotation rate of 2100 rpm,
it died out when going to higher value. 
Here, we expect that some cavitation has occurred.
During run 1, however, it was possible 
to study the dependence of the saturation level on the
rotation rate at rather fine gradation.

\subsection{July 2003}

\begin{figure}
\begin{center}
\epsfxsize=13.4cm\epsfbox{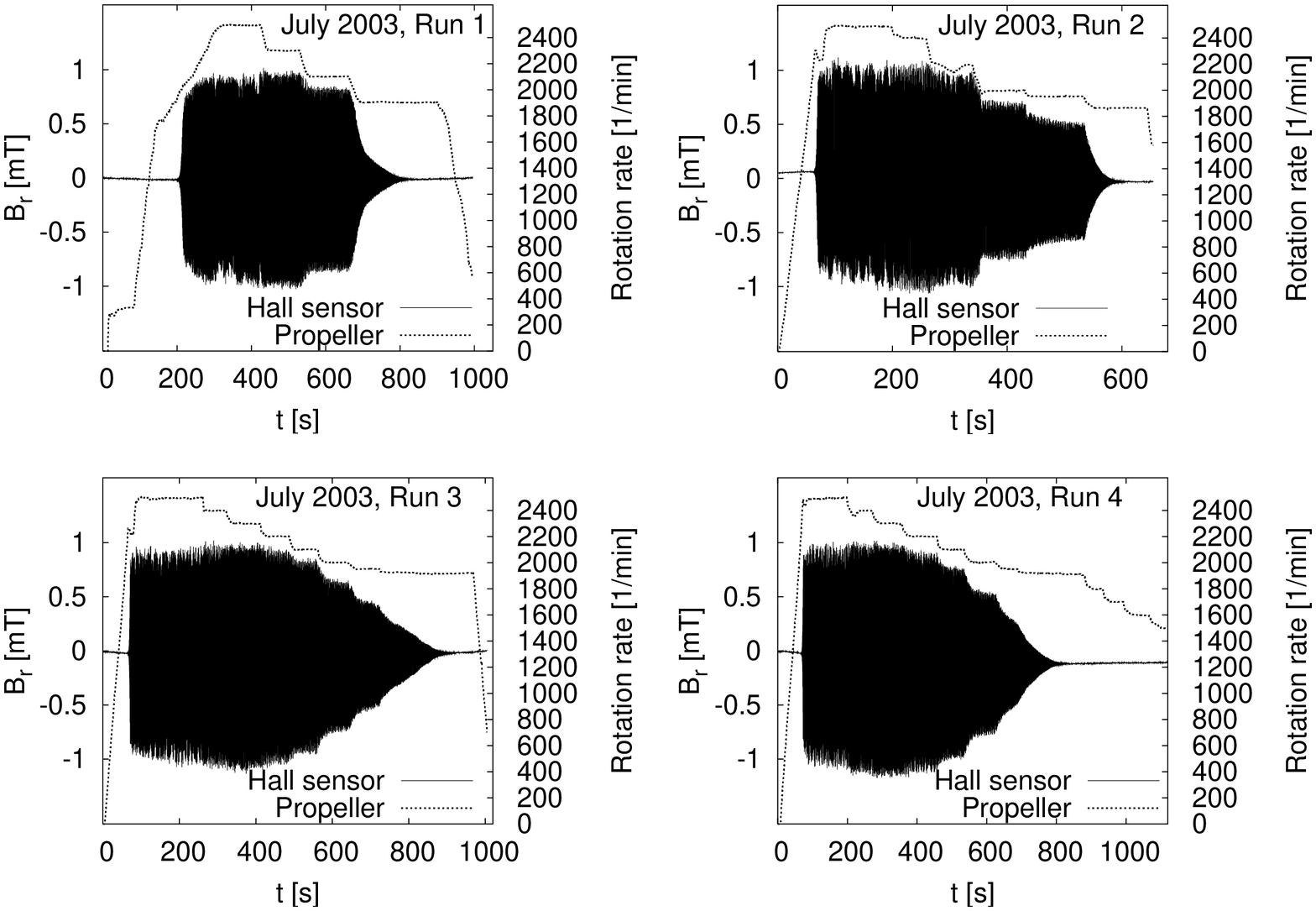}
\end{center}
\caption{Experiments in July 2003. Note the very slow decay of the field
when the rotation rate was reduced to a slightly subcritical values.}
\label{fig7}
\end{figure}

The July 2003 campaign has again delivered four ''perfect'' runs 
(figure \ref{fig7}). Most impressive here are the very
slow decays of the field after the critical rotation rate was
reduced below the critical values. 

\subsection{May 2004}

\begin{figure}
\begin{center}
\epsfxsize=13.4cm\epsfbox{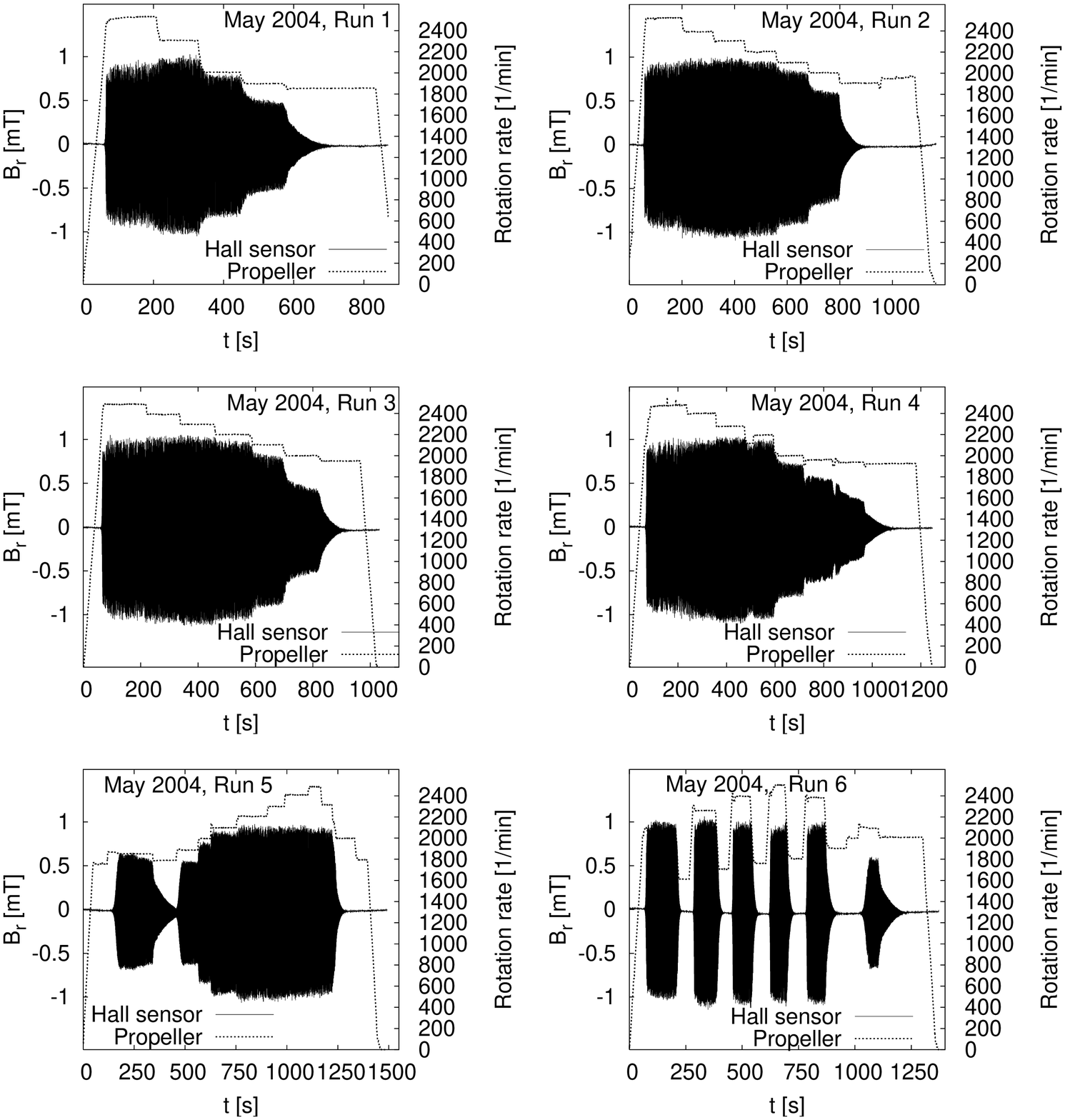}
\end{center}
\caption{Experiments in May 2004. Note the multiple 
switching on/off of the dynamo for Run 6.}
\label{fig8}
\end{figure}

A further enlargement of the database was obtained 
with the 6 successful runs of May 2004 (figure \ref{fig8}). 
Run 6 shows how
the dynamo is switched on and off 6 times simply by 
crossing the critical rotation rate from below or from above.

A novelty of the May 2004 campaign was the 
measurement of pressure data in the inner dynamo channel
by a piezoelectric sensor that was flash mounted at the 
innermost wall. As will be shown below, 
these measurements provided interesting hydrodynamic data
for the characterization of the turbulence.

\subsection{February/March 2005}

\begin{figure}
\begin{center}
\epsfxsize=13.4cm\epsfbox{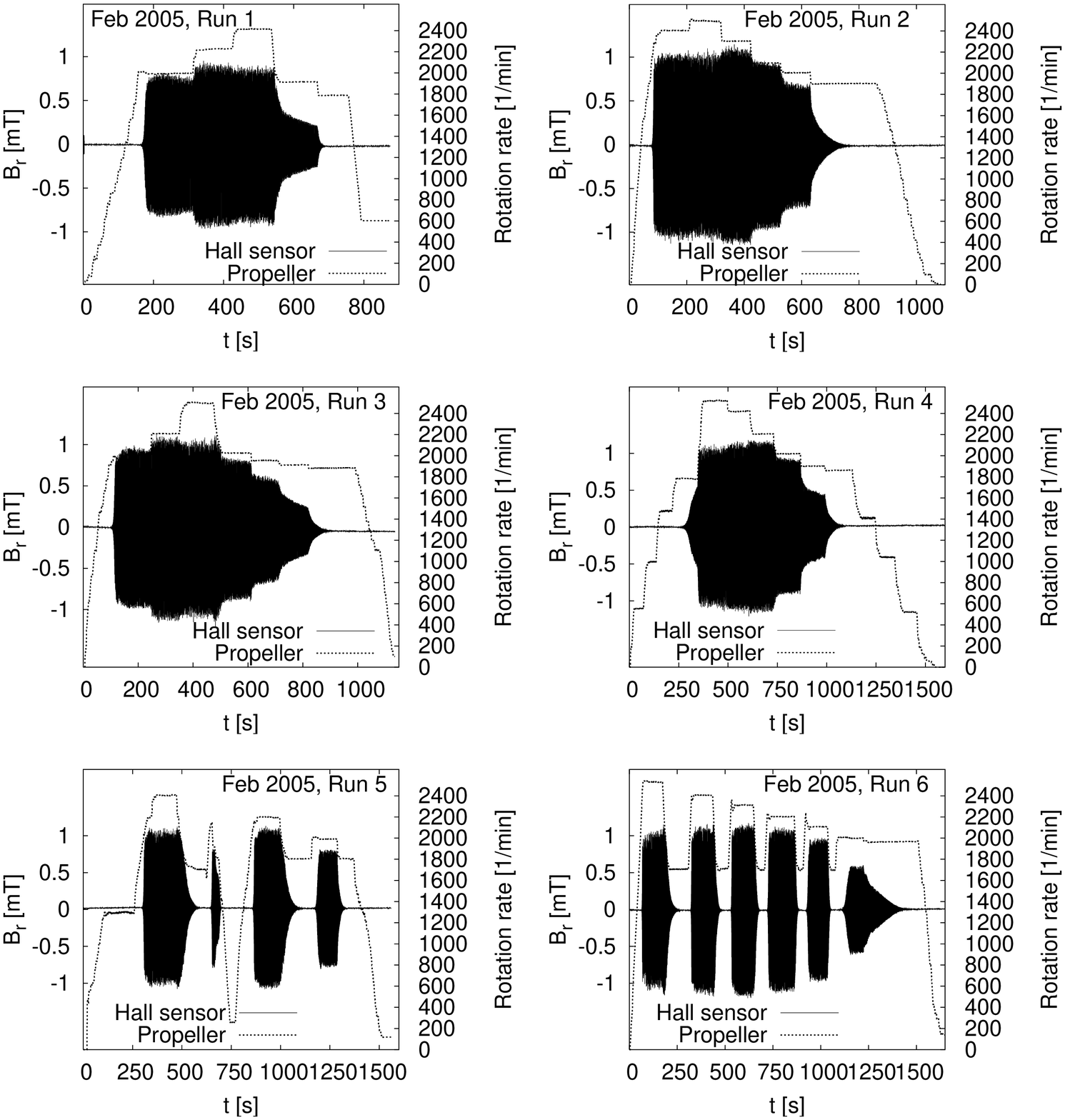}
\end{center}
\caption{Experiments in February/March 2005. Note the multiple 
switching on/off of the dynamo during runs 5 and 6.}
\label{fig9}
\end{figure}

Quite as successful as the runs in May 2004 were the runs in
February/March 2005 (figure \ref{fig9}). 
What was new in this campaign was the installation
of two traversing rails with Hall sensors moving 
in axial direction outside the dynamo and induction 
coils moving in
radial direction within the dynamo module. These 
traversing sensor rails allowed for the detailed 
determination of the
spatial structure of the magnetic eigenfield.

\subsection{July 2007}

\begin{figure}
\begin{center}
\epsfxsize=13.4cm\epsfbox{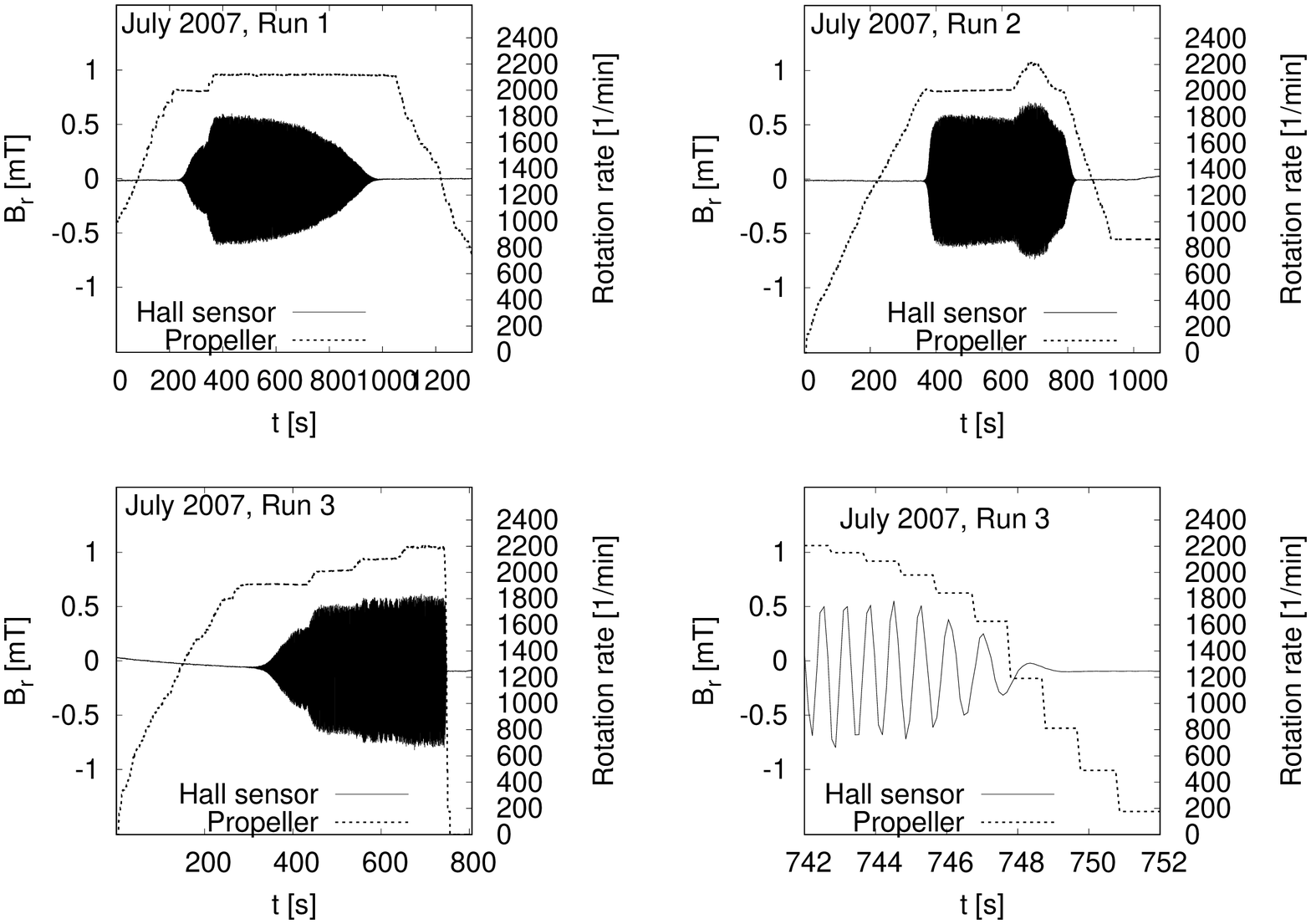}
\end{center}
\caption{Experiments in July 2007. For unknown reasons, 
the dynamo capability was already reduced. Run 3 terminated with
a breaking belt, leading to a sudden decay of the propeller rotation rate
and a quick dying out of the dynamo (lower right panel).}
\label{fig10}
\end{figure}

July 2007 marks the end of the first successful 
experimental series at the Riga dynamo facility
(figure \ref{fig10}). For 
unknown reasons (oxides, cavitation...?)
the dynamo capability was already slightly reduced.
We will see further below 
that also the growth rate was markedly reduced.
Yet, run 1 shows an interesting feature: 
while keeping the rotation rate constant for appr. 11
minutes, the dynamo dies out due to the fact that the
electrical conductivity decreases with the 
slowly increasing temperature of the liquid sodium.

The campaign terminated with a breaking of one 
of the rubber 
belts connecting the motors with the propeller shaft. 
The details of this dramatic 
event, which occurred at the end of  run 3, 
are documented in the lower right panel of 
figure \ref{fig10}.

\section{Main results}

In this section we will summarize some of the most 
important results of the dynamo experiments, 
with a strong focus
on the comparison with numerical predictions. 

\subsection{Growth rates and frequencies}

The series of experiments comprised a wealth
of time periods in which the velocity can be considered 
as stationary. This either means that the magnetic field was
low enough not to disturb the original flow (kinematic regime), 
or that a certain saturation level of the magnetic field 
was maintained for a longer period (saturated regime).

\begin{figure*}
\begin{center}
\epsfxsize=10cm\epsfbox{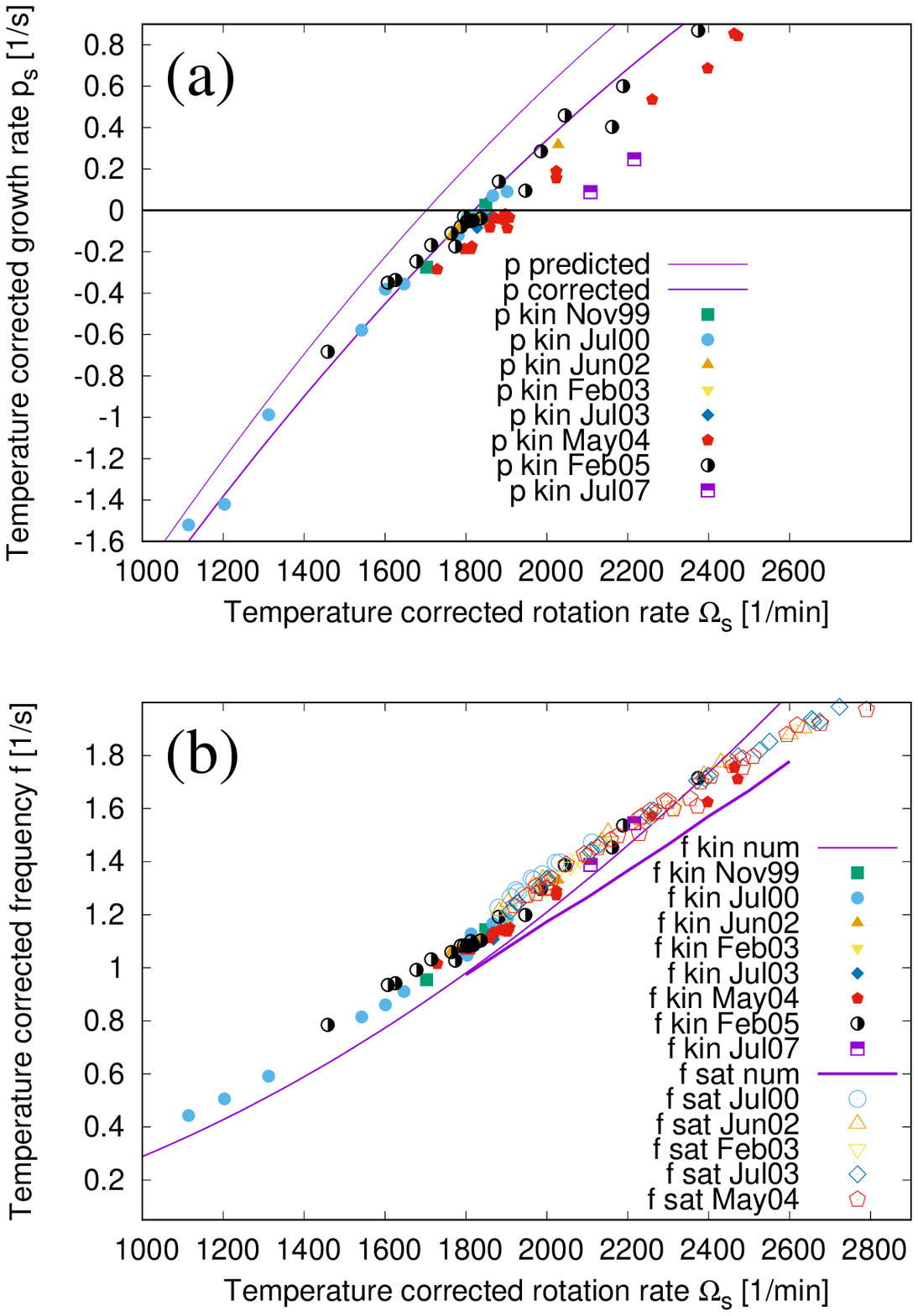}
\end{center}
\caption{Growth rates (a) and frequencies (b) 
in the kinematic and the saturated
regime for the various campaigns. The 
numerical predictions for
the kinematic case are based on the 2D code 
described in the text. For the
corrected curves, the effect of the finite wall 
thickness (determined with a
1D code) was additionally taken into account. All
rotation rates, growth rates and frequencies were
re-scaled to a common reference 
temperature of 157$^{\circ}$C. Note the significant reduction
of the growth rates for the two ``bad'' runs from July 2007.
}
\label{fig11}
\end{figure*}

In figure \ref{fig11} we put together several numerical 
predictions
and a rather complete set of the measured values for the growth
rate $p$ and the frequency $f$. The numerically 
computed growth rates comes in two
versions: "p predicted" is the result of the 2D code 
(as discussed above),  
relying on the velocity measurements in water which were
carried out before the sodium experiment 
(Christen, H\"anerl \& Will 1998; 
Stefani, Gerbeth \& Gailitis  1999). "p corrected" 
represents a slight correction of "p predicted", for which
the lower conductivity of the stainless steel walls has 
been incorporated in
the 1D-code, and the obtained corrections were then added 
to the 2D results (admittedly, a somewhat ''hybrid'' 
method which seems, however, quite
reliable). The corresponding two predictions for the frequencies 
do barely differ and are therefore summarized as one single curve 
"f kin num". The second prediction for the frequencies, "f sat num",
concerns the values in the saturated regime. It has been
obtained by solving equation (\ref{eq5}) simultaneously with
the induction equation, and inferring the frequency when the system
has relaxed into saturation.

The validity of this saturation  model (which gives 
automatically a zero growth rate) can be judged from 
the dependence
of the resulting eigenfrequency in figure \ref{fig11}. 
We observe here a quite reasonable correspondence with the 
measured data, in particular with view on 
the slightly declined 
slope of the curve. A minor jump of the measured 
eigenfrequencies between the kinematic and the saturated 
regime might be attributed to the arising 
fluid rotation in the outermost cylinder, which
is not  incorporated in our simple back-reaction model.

\subsection{Power increase}

The excess power that is necessary to overcome 
the Ohmic losses
due to the self-excited magnetic field is one of the
most important features in the saturation regime of 
a dynamo. Unfortunately, its precise determination is 
far from trivial. One reason is that in the saturation regime it is
barely possible to measure the motor power in the purely 
hydraulic case, without any magnetic field back-reaction, 
although the power in the purely hydraulic case
scales over a wide range of rotation rates with the cube 
of the rotation rate. 
Yet, slightly above the critical rotation rate, 
where the magnetic field growth rate is slow,
there are time periods long enough to 
determine the purely hydraulic power before
the magnetic field has reached large values.

Another point is that, in particular
for high rotation rates, the dynamo behaviour was 
not completely reproducible
due to different Argon pressure regimes resulting in different couplings 
of the sodium flow to the propeller rotation. This means that the few
available hydraulic
power values in the high rotation rate region cannot be taken as an
accurate  reference 
power to compare with the full power under the action of saturated
Lorentz forces. Another minor point is connected with the not perfect
reproducibility of the runs, meaning that the critical rotation rate is
not exactly the same for all runs. A further issue is 
the effect of the temperature on the
conductivity.

\begin{figure*}
\begin{center}
\epsfxsize=10cm\epsfbox{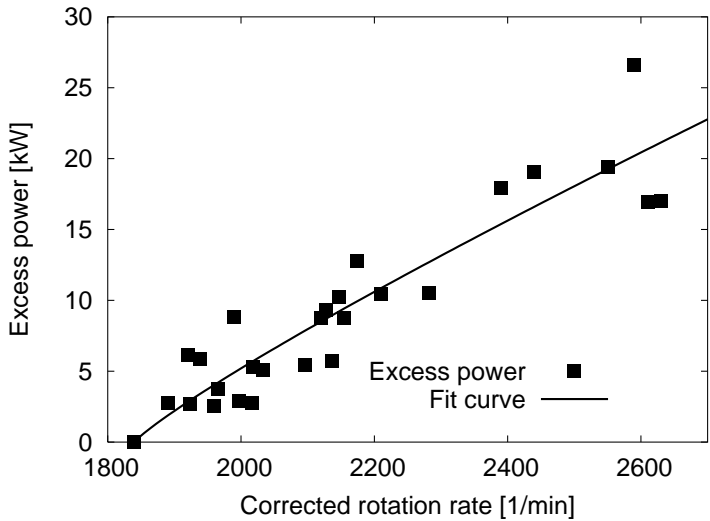}
\end{center}
\caption{Excess motor power due to the action of the Lorentz force,
estimated by comparing the total power with the
(not very safely determinable) pure hydraulic power.
}
\label{fig12}
\end{figure*}

With all those uncertainties in mind, and 
trying to correct for all possible
inconsistencies, we inferred the dependence 
of the excess power on the difference of 
the rotation rate and the critical rotation
rate. Figure \ref{fig12} shows the corresponding data.
Compared with the purely hydraulic power
in the order of 200 kW, the typical 20 kW represent 
only a 10 per cent increase. This rather 
flat increase reflects the {\it fluid character} 
of the dynamo,
in which the sodium flow evades and
deforms under the influence of the Lorentz forces, 
while the resulting deterioration of the dynamo condition 
makes the growth rate drop down to zero.

\subsection{Radial field distribution}

\begin{figure*}
\begin{center}
\epsfxsize=10cm\epsfbox{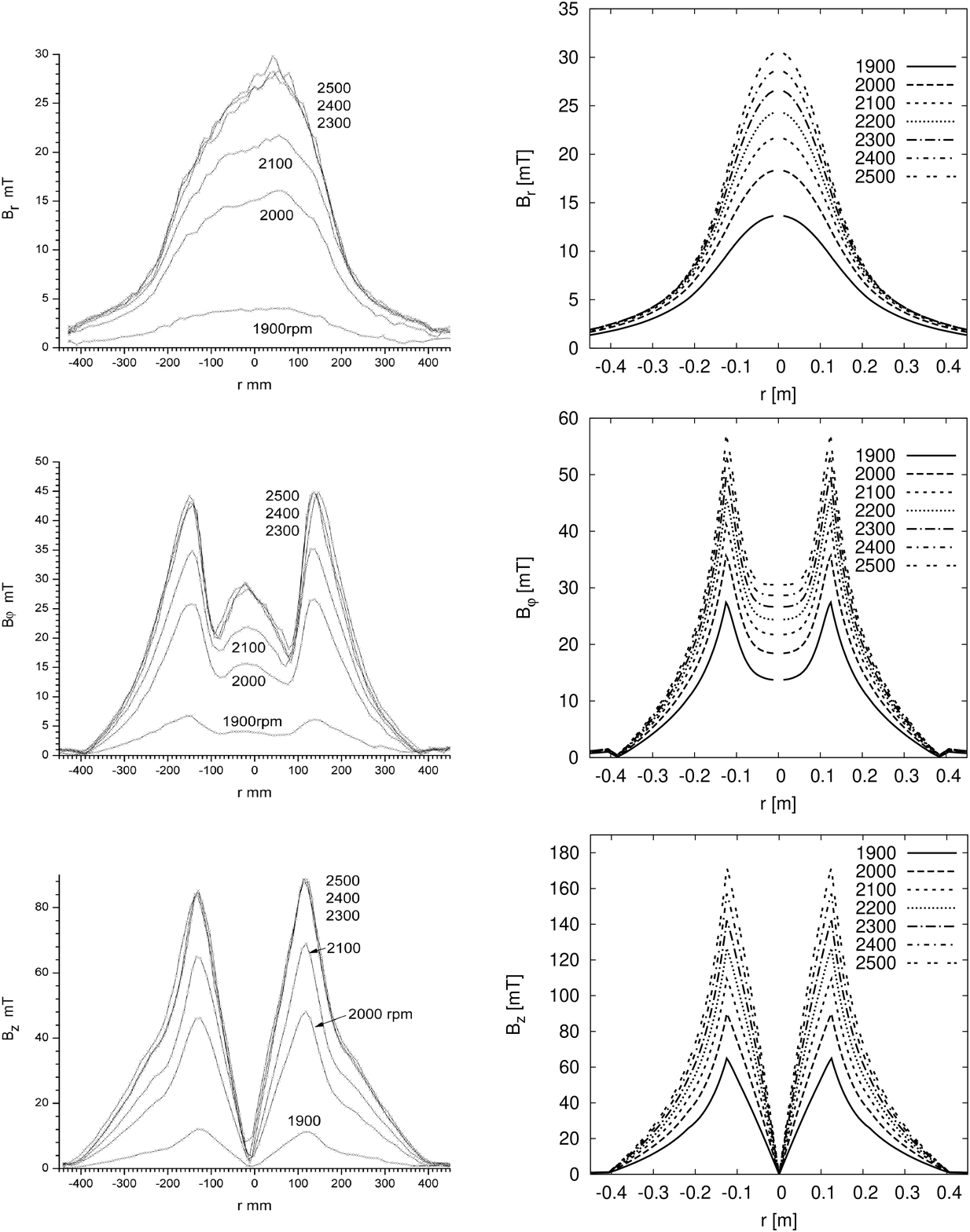}
\end{center}
\caption{Radial dependencies of the three field components 
measured at the lower port in Run 2 from Feb 2005 (left),
and corresponding numerical predictions (right).
}
\label{fig13}
\end{figure*}

The newly installed lances that were radially and axially 
movable through and along the dynamo allowed to determine the radial and
axial field dependencies in much detail. Figure   \ref{fig13}
shows the dependence of the three field components 
on the radius, as measured by the radial lance in the lower port. 
The left column shows the measured values,
the right column shows the corresponding numerical predictions.
While a perfect quantitative agreement cannot be expected, 
a nice agreement of the form of the curves 
is generally observed. A remarkable deviation becomes 
visible, however,
for the $B_{\phi}(r)$ profiles at higher rotation rate, which
develop a sort of ''bump'' at $r=0$ which is not reflected by the 
numerical value.

\subsection{Axial field distribution}
\begin{figure*}
\begin{center}
\epsfxsize=13.4cm\epsfbox{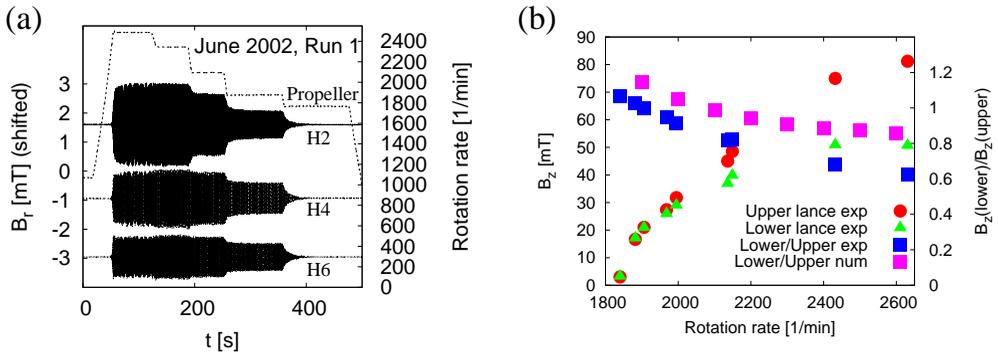}
\end{center}
\caption{Upward shift of the field pattern in the saturated 
regime. (a) Run 1 from June 2002 (see figure 5), enhanced by 
the signals of H2 and H6. The amplitude ratios between the
upper sensor H2 and the lower sensors H4 and H6  
increase markedly for higher rotation rates (for the sake of
better visibility, the signals are shifted along the ordinate 
axis).
(b) Measured values and decreasing ratio of axial fields  
at the lower and upper lances. The numerical prediction for the ratio 
shows the same
decreasing trend, although not as strong as the measured values.
}
\label{fig14}
\end{figure*}

While the radial field distribution has turned 
out not to change greatly from the 
kinematic regime to the saturation regime,
the axial field is affected significantly  by the back-reaction.
The braking of the azimuthal velocity component, 
which accumulates downstream from the propeller, results in a
deteriorated self-excitation capability of the flow in the 
lower parts of the dynamo and therefore in an upward shift
of the entire magnetic field structure. This is 
illustrated in figure \ref{fig14}. First, figure \ref{fig14}a
reiterates run 1 from June 2002 (as in figure 5), 
enhanced now by the signals of the sensors 
H2 and H6. Evidently, the ratios between the amplitudes
of the upper sensor H2 and the lower sensors 
H4 and H6 grow with increasing rotation rates.
Figure \ref{fig14}b shows the measured axial magnetic 
field amplitudes
(from the same June 2002 campaign) at the upper and lower 
ports and their ratio in dependence on the rotation rate.
The typical decrease of the ratio of lower to upper 
fields is also reflected in the numerical model.

\subsection{Turbulence properties and magnetic field spectra}

\begin{figure*}
\begin{center}
\epsfxsize=10cm\epsfbox{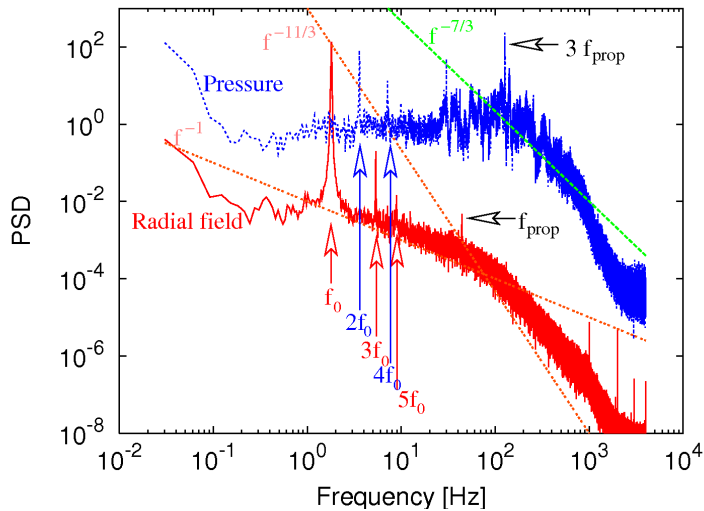}
\end{center}
\caption{Spectra of the radial magnetic field and the pressure, both measured on
the lower level in the inner cylinder. The pressure sensor was flash mounted to the wall.
The data are for a rotation rate of 2530 rpm.
The peaks are multiples of the eigenfrequency $f_0$ of the magnetic field. 
There are also peaks at the motor
rotation rate and its triple value (because of three wings).}
\label{fig15}
\end{figure*}

Apart from the back-reaction effects on the large scale 
flow structure, the effect on the fluctuations is also important.
In figure \ref{fig15} two spectra are shown, one from a Hall sensor
which is located on the lower measurement level in the central cylinder,
about 
2 cm from the wall. The other spectrum results from the data of the 
mentioned pressure sensor, which is also mounted on the lower level.
The data were recorded during run 1 of the May 2004 campaign.

Not surprisingly, the key feature of the magnetic spectrum is
its peak at the
eigenfrequency $f_0$. Yet, there are further peaks at the triple 
and at the five-fold frequency.
Neither of these peaks is seen in the pressure spectrum. 
Here, we detect 
a dominant peak at  $2 f_0$ and some smaller peak at the
$4 f_0$. Obviously, the  
magnetic eigenfield with the angular mode $m=1$ produces a Lorentz force
with a dominant $m=0$ part, but also with an $m=2$ part. The latter 
part influences the velocity and is also mirrored by the pressure 
peak at $2 f_0$. Then, this $m=2$ mode of the velocity 
induces, together with the dominant magnetic $m=1$ mode,
a new contribution
with $m=3$ in the magnetic field. The product of $m=1$ and $m=3$ modes of
the magnetic field produces the $m=4$ mode in the pressure. 
All the arguments for 
$m$ transfer to the multiples of the 
frequency since the measurement is done at a fixed position.

Concerning the inertial range of the spectrum, we have plotted
a typical $f^{-11/3}$ law for the magnetic field in the inertial range 
and a  $f^{-7/3}$ law for the pressure   
for comparison, without claiming perfect coincidences.
Between the main field frequency $f$ and the propeller 
frequency $f_{prop}$ 
there seems to be a region with a $f^{-1}$ scaling.

\begin{figure*}
\begin{center}
\epsfxsize=10cm\epsfbox{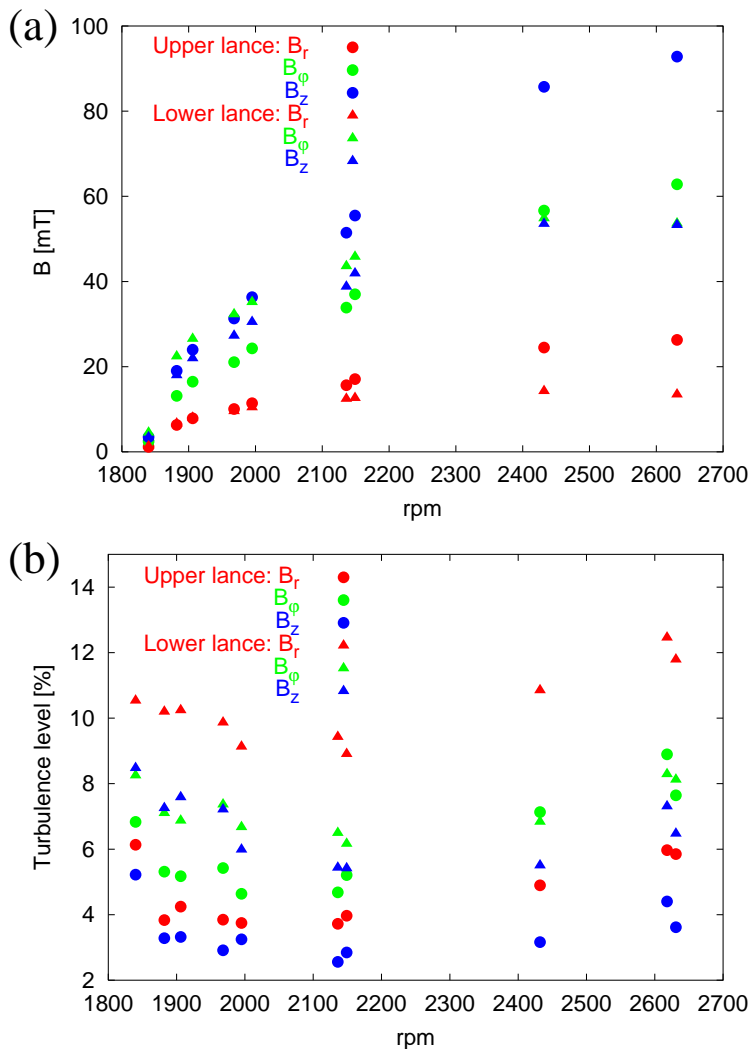}
\end{center}
\caption{Amplitudes (a) and turbulence levels (b) 
measured in the middle of the
back-flow cylinder at the lower and upper port
}
\label{fig16}
\end{figure*}

Another interesting result concerns the turbulence level.
For this purpose we have analyzed the data from the June 2002 campaign.
Specifically, we have used the data from the Hall sensors positioned 
at the upper and lower port, in the middle of the back-flow 
region. The sensors measure a dominant sine signal from
the magnetic eigenfield whose amplitude is shown in 
figure \ref{fig16}a. The deviation from the dominant
sine signal is then attributed to the flow turbulence,
whose level (in per cent) is given in 
figure \ref{fig16}b. The first observation to make here is
the reduced turbulence level at the upper 
sensors which might be due to some calming of the 
flow when it goes upward. The second, and more interesting observation 
is the general minimum of the turbulence level around 2100 rpm,
for which we have no explanation up to present.

\section{Synopsis, conclusions, prospects}

Let us summarize the main results of the Riga dynamo experiments:
In the kinematic regime the measured growth rates, frequencies 
and the spatial field structure were in close
correspondence with the numerical predictions of a 2D 
code, in particular, when a correction for the resistive effects 
of the inner stainless steel walls was made. 
This is, actually, not a trivial point. Before the 
experiment was really performed, some unrecognized 
influence of the (low level) turbulence 
on the self-excitation threshold, or a greater 
sensitivity to the precise boundary and interface 
conditions, could not be ruled out entirely. Now,
we can conclude that the ``elementary cell'' of a 
dynamo works robustly in the laboratory, and that its 
kinematic behaviour is predictable with an error 
margin of a few per cent.

In contrast to that, the understanding of the saturation 
regime is a much more involved. We have shown that a rather
simple back-reaction model, which computes the selective 
breaking of the azimuthal flow component along the 
streamlines, comes close to explaining the observed 
back-reaction effect, such as the continuing increase of the
frequency and the upward shift of the field pattern. 
However, there are features which are not explained by this 
simple model: the developing maximum of 
$B_{\varphi}$ at $r=0$ (see figure \ref{fig13}) is just a 
case in point. A further yet unexplained feature is the
minimum of the turbulence level for some medium 
degree of supercriticality, as evidenced in figure 
\ref{fig16}.

After a complete disassembly, reconstruction, and 
refurbishment
the Riga dynamo was started again with short
campaigns in June 2016 and April 2017. 
Part of the results were published by 
Gailitis and Lipsbergs (2017), and further 
details will be discussed elsewhere. 
If the attainment of the old
performance parameters can be confirmed, 
a significant modification of the flow
structure is planned for the future.
This is motivated by the numerical finding  that
a specific ''de-optimization'' of the flow
field in the Riga dynamo could lead to
a vacillation between two states 
with different kinetic and magnetic energies 
(Stefani, Gailitis \& Gerbeth 2011).

In concert with the complementary and nearly 
contemporaneous Karlsruhe experiment, the Riga 
dynamo experiment has pioneered a couple of 
further activities for studying dynamo action, and 
related magnetic instabilities, in the
liquid metal lab. Important new results were, 
among others, 
the observation of magnetic field reversals
in the French VKS experiment (Berhanu {\it et al.} 2010), 
and
the experimental demonstration of the helical 
(Stefani {\it et al.}  2006)
and azimuthal (Seilmayer {\it et al.} 2014) 
magnetorotational instability as well as the 
current-driven Tayler instability (Seilmayer {\it et al.} 2012).
A comparative survey of these, and further dynamo
related experiments, can be found in
the review papers by Stefani, Gailitis \& Gerbeth 
(2008) and Stefani {\it et al. (2017).


\begin{thebibliography}{99}

\bibitem[Berhanu et al.(2007)]{BERHANU} 
{\sc Berhanu, M. et al.} 2007
Magnetic field reversals in an experimental turbulent dynamo.
{\em EPL} {\bf 77}, 59001.


\bibitem[Christen, H\"anel \& Will(1998)]{CHRISTEN}
{\sc Christen,M., H\"anel, H. \& Will, G.} 1998 Entwicklung
der Pumpe f\"ur den hydrodynamischen Kreislauf des Rigaer
''Zylinderexperimentes. In {\it{Beitr\"age zu
Fluidenergiemaschinen 4}} (ed. W.H. Faragallah \& G. Grabow)
pp. 111-119, 
Faragallah-Verlag und Bildarchiv, Sulzbach/Ts.



\bibitem[Gailitis(1967)]{GAIL67} 
{\sc Gailitis, A.} 1967
Self-excitation conditions for a 
laboratory model of a geomagnetic dynamo.
{\em Magnetohydrodynamics}
{\bf {3}}, No. 3, 23--29.

\bibitem[Gailitis \& Freibergs(1976)]{GAFR76}
{\sc Gailitis, A. \& Freibergs, Ya.} 1976
Theory of a helical MHD dynamo.
{\em Magnetohydrodynamics}
 {\bf {12}}, 127--129.

\bibitem[Gailitis \& Freibergs(1980)]{GAFR80}
{\sc Gailitis, A. \& Freibergs, Ya.} 1980 
Nonuniform model of a helical dynamo.
{\em Magnetohydrodynamics} {\bf {16}}, 116--121.

\bibitem[Gailitis et al.(1987)]{GAI87}
{\sc Gailitis, A. Karasev, B.G., Kirillov, I.R.,
Lielausis, O.A., Luzhanskii, S.M., Ogorodnikov, A.P.
\& Preslitskii, G.V.} 1987
Experiment with a liquid-metal model of an MHD dynamo.
{\em Magnetohydrodynamics} {\bf 23}, 349--353.

\bibitem[Gailitis(1996)]{GAI96}
{\sc Gailitis, A.} 1996
Design of a liquid sodium MHD dynamo experiment.
{\em Magnetohydrodynamics} {\bf 32}, 58--62.


\bibitem[Gailitis et al.(2000)]{PRL1}
{\sc Gailitis, A., Lielausis, O., Dement'ev, S.,  Platacis, E., Cifersons, A.,
Gerbeth, G., Gundrum, T., Stefani, F., Christen, M., H\"anel, H. \&  Will, G.} 2000
Detection of a flow induced magnetic field eigenmode in the Riga dynamo facility.
{\em Phys. Rev. Lett.} {\bf 84}, 4365--4368.


\bibitem[Gailitis et al.(2001)]{PRL2}
{\sc Gailitis, A., Lielausis, O., 
Platacis, E., Dement'ev, S., Cifersons, A.,
Gerbeth, G., Gundrum, T., Stefani, F., Christen, M., \&  Will, G.} 2001
Magnetic field saturation in the Riga dynamo experiment.
{\em Phys. Rev. Lett.} {\bf 86}, 3024--3027.

\bibitem[Gailitis et al.(2001a)]{MAHYD1}
{\sc Gailitis, A., Lielausis, O., Platacis, E.,
Gerbeth, G. \& Stefani, F.} 2001
On the results of the Riga dynamo experiments.
{\em Magnetohydrodynamics} {\bf 37}, 71--79.

\bibitem[Gailitis et al.(2002)]{RMP}
{\sc Gailitis, A., Lielausis, O., Platacis, E.,
Gerbeth, G. \& Stefani, F.} 2002
Laboratory experiments on hydromagnetic dynamos.
{\em Rev. Mod. Phys.}  {\bf 74},  973-990.

\bibitem[Gailitis et al.(2002a)]{MAHYD2}
{\sc Gailitis, A., Lielausis, O., 
Platacis, E., Dement'ev, S., Cifersons, A.,
Gerbeth, G., Gundrum, T., Stefani, F., Christen, M., \&  Will, G.} 2002
Dynamo experiments at the Riga sodium facility. 
{\em Magnetohydrodynamics} {\bf 38}, 5-14.

\bibitem[Gailitis et al.(2002b)]{MAHYD3}
{\sc Gailitis, A., Lielausis, O., Platacis, E.
Gerbeth, G. \& Stefani, F.} 2002
On back-reaction effects in the Riga dynamo experiment.
{\em Magnetohydrodynamics} {\bf 38}, 15-26.

\bibitem[Gailitis et al.(2003)]{SURV} 
{\sc Gailitis, A., Lielausis, O., Platacis, E.:
Gerbeth, G. \& Stefani, F.} 2003 
The Riga dynamo experiment.
{\em Surv. Geopyhs.} {\bf 24}, 247-267.

\bibitem[Gailitis et al.(2004)]{PLASMA} 
{\sc Gailitis, A., Lielausis, O., Platacis, E.,
Gerbeth, G. \& Stefani, F.} 2004
Riga dynamo experiment and its theoretical background.
{\em Phys. Plasmas} {\bf 11}, 2838-2843.

\bibitem[Gailitis et al.(2008)]{CRAS} 
{\sc Gailitis, A., Gerbeth, G.,  Gundrum, Th., 
Lielausis, O., Platacis, E. \& Stefani, F.} 2008 
History and results of the Riga dynamo experiments.
{\em C.R. Phys.} {\bf 9}, 721-728. 

\bibitem[Gailitis \& Lipsbergs(2017)]{GUNTIS}
{\sc Gailitis, A. \& Lipsbergs, G.} 2017 
2016 year experiments at Riga dynamo facility,
{\em Magnetohydrodynamics}  {\bf 53}, 349-356.

\bibitem[Kenjere\v{s} et al.(2006)]{SASA1}
{\sc Kenjere\v{s}, S., Hanjali\'{c}, K., Renaudier, S., 
Stefani, F., Gerbeth, G. \& Gailitis, A.} 2006
Coupled fluid-flow and magnetic-field 
simulation of the Riga dynamo experiment.
{\em Phys. Plasmas} {\bf 13}, 122308.


\bibitem[Kenjere\v{s} \& Hanjali\'{c}(2007)]{SASA2}
{\sc S. Kenjere\v{s}, S. \& Hanjali\'{c}, K.} 2007
Numerical simulation of a turbulent magnetic dynamo.
{\em Phys. Rev Lett.} {\bf 98}, 104501. 



\bibitem[M\"uller \& Stieglitz(2002)]{MUST2} 
{\sc M\"uller, U. \&  Stieglitz, R.} 2002
The Karlsruhe dynamo experiment.
{\em Nonl. Proc. Geophys.} {\bf 9}, 165-170.


\bibitem[M\"uller, Stieglitz \& Horanyi(2004)]{MUST3} 
{\sc M\"uller, U.,  Stieglitz, R. \& Horanyi, S.} 2004 
A two-scale hydromagnetic dynamo experiment.
{\em J. Fluid Mech}. {\bf 498}, 31-71.

\bibitem[M\"uller, Stieglitz \& Horanyi(2006)]{MUST4} 
{\sc M\"uller, U.,  Stieglitz, R. \& Horanyi, S.} 2006  
Complementary experiments at the Karlsruhe dynamo test facility.
{\em J. Fluid Mech.} {\bf 552}, 419-440.


\bibitem[Steenbeck, Krause \& R\"adler(1966)]{STEENBECKKRAUSERAEDLER}
{\sc Steenbeck, M., Krause, F. \& R\"adler, K.-H.} 1966
Berechnung der mittleren Lorentz-Feldst\"arke 
$\overline{{\bf v} \times {\bf b}}$ f\"ur ein 
elektrisch leitendendes Medium in turbulenter, 
durch Coriolis-Kr\"afte beeinflu\ss ter Bewegung.
{\em Z. Naturforsch.} {\bf 21a}, 369--376.


\bibitem[Steenbeck et al.(1967)]{STEENBECKALPHA}
{\sc Steenbeck, M., Kirko, I.M., Gailitis, A., Klawina, A.P.,
Krause, F., Laumanis, I.J. \& Lielausis, O.A.} 1967
Der experimentelle Nachweis einer elektromtorischen 
Kraft l\"angs eines \"au\ss eren Magnetfeldes, induziert durch die
Str\"omung fl\"ussigen Metalls ($\alpha$-Effekt).
{\em Mber. Dt. Ak. Wiss.} {\bf 9}, 714--719.

\bibitem[Seilmayer et al.(2012)]{Seilmayer2012} 
{\sc Seilmayer, M. \etal} 2012 
Experimental evidence for a transient 
Tayler instability in a cylindrical liquid-metal column.
{\em Phys. Rev. Lett.} {\bf 108}, 244501.

\bibitem[Seilmayer et al.(2014)]{Seilmayer2014} 
{\sc Seilmayer, M. \etal} 2014
Experimental evidence for nonaxisymmetric 
magnetorotational instability in a rotating 
liquid metal exposed to an azimuthal magnetic field.
{\em Phys. Rev. Lett.} {\bf 113}, 024505.


\bibitem[Stefani, Gailitis \& Gerbeth(1999)]{KLUWER}
{\sc Stefani, F., Gerbeth, G. \& Gailitis, A.} 1999
Velocity profile optimization for the Riga dynamo experiment.
In {\it{Transfer
Phenomena in Magnetohydrodynamic and Electroconducting Flows}},
(ed. A. Alemany, Ph. Marty \& J. P. Thibault),
pp. 31-44, Kluwer.

\bibitem[Stefani et al.(2006)]{Stefani2006} 
{\sc Stefani, F. \etal} 2006
Experimental evidence for magnetorotational 
instability in a Taylor-Couette flow under 
the influence of a helical magnetic field.
{\em Phys. Rev. Lett.} {\bf 97}, 184502.

\bibitem[Stefani, Gailitis \& Gerbeth(2008)]{ZAMM}
{\sc Stefani, F., Gailitis, A., Gerbeth, G.} 2008
Magnetohydrodynamic 
experiments on cosmic magnetic fields. 
{\em ZAMM} {\bf 88}, 930-954.

\bibitem[Stefani, Giesecke \& Gerbeth(2009)]{TCFD}
{\sc Stefani, F., Giesecke, A., Gerbeth, G.} 2009
Numerical simulations of liquid metal experiments 
on cosmic magnetic fields.
{\em Theor. Comp. Fluid Dyn.} {\bf 23}, 405-429.

\bibitem[Stefani, Gailitis \& Gerbeth(2011)]{AN}
{\sc Stefani, F., Gailitis, A. \&  Gerbeth, G.} 2011
Energy oscillations and a possible route to chaos in a modified Riga dynamo.
{\em  Astron. Nachr.} {\bf 332}, 4--10.

\bibitem[Stefani et al. (2017)]{Stefani2017} 
{\sc Stefani, F. \etal} 2017
Magnetic field dynamos and magnetically triggered flow instabilities.
{\em IOP Conf. Ser.: Mater. Sci. Eng.} {\bf 228}, 012002.


\bibitem[Stieglitz \& M\"uller(2001)]{STMU}
{\sc Stieglitz, R., M\"uller, U} 2001
Experimental demonstration of a homogeneous two- scale dynamo.
{\em Phys. Fluids} {\bf 13}, 561--564.

\end{thebibliography}
\end{document}